\documentclass[aps,prd,
groupedaddress,amssymb,
showpacs,nofootinbib]{revtex4}

\usepackage{amsmath,amssymb}

\usepackage[usenames]{color}

\usepackage{graphicx}
\usepackage{mathptmx}
\usepackage{bbm}
\usepackage{bm}
\usepackage{times}

\newcommand{\beq}{\begin{eqnarray}}
\newcommand{\eeq}{\end{eqnarray}}

\begin{document}

\onecolumngrid



\title{Summing Planar Diagrams by an Integrable Bootstrap II}



\author{Peter \surname{Orland}}

\email{orland@nbi.dk}


\affiliation{1. Baruch College, The 
City University of New York, 17 Lexington Avenue, 
New 
York, NY 10010, U.S.A. }

\affiliation{2. The Graduate School and University Center, The City University of New York, 365 Fifth Avenue,
New York, NY 10016, U.S.A.}



\begin{abstract}

We continue our investigation of correlation functions of the large-$N$ (planar) limit of the 
$(1+1)$-dimensional principal chiral sigma model, whose bare field 
$U(x)$ lies in the fundamental matrix representation of 
${\rm SU}(N)$. We find all the form factors of the 
renormalized field $\Phi(x)$. An exact formula for Wightman and time-ordered correlation functions is found.

\end{abstract}


\maketitle

\section{Introduction}
\setcounter{equation}{0}
\renewcommand{\theequation}{I.\arabic{equation}}

In Reference \cite{SU-infty-FF} (henceforth referred to as {\bf I}), we found some form factors of the renormalized field $\Phi$, of the $(1+1)$-dimensional principal chiral model in 't~Hooft's planar limit 
\cite{'t}. In particular, we obtained the one- and three-excitation form factors 
(the two-excitation form factor is zero). These form factors
yield expressions for the correlation functions of the renormalized field, for large separations. In this paper we extend our results to all form factors
of $\Phi$. We thereby obtain exact expressions for 
correlation functions. Thus the planar diagrams are {\em completely} resummed.

Our technique is a combination of the form-factor axioms \cite{Smirnov} and the $1/N$-expansion of the exact {\em S} matrix \cite{AALS}, \cite{Wiegmann}, \cite{Pol-Wieg}. A related development is the determination of the $1/N$-expansion of the two- and four-excitation 
form factors of current operators, by A. Cort\'es Cubero \cite{axel}. 

We do not assume the reader is well-versed in form-factor lore, but take for granted 
acquaintance with integrability and 
two-dimensional {\em S}-matrix theory. 

The bare 
field of the principal chiral model is a matrix $U(x) \in {\rm SU}(N)$, $N\ge 2$, where $x^{0}$ and 
$x^{1}$ are the time and space coordinates, respectively. The action is
\beq
S=\frac{N}{2g_{0}^{2}}\int d^{2}x \;\eta^{\mu\nu}\;{\rm Tr}\,\partial_{\mu}U(x)^{\dagger}\partial_{\nu}
U(x),
\label{action}
\eeq
where $\mu, \nu=0,1$, $\eta^{00}=1$, $\eta^{11}=-1$, $\eta^{01}=\eta^{10}=0$, where $g_{0}$ is the coupling, This action is invariant under the global transformation 
$U(x)\rightarrow V_{L}U(x)V_{R}$,  for two constant matrices $V_{L}, \,V_{R}\in {\rm SU}(N)$. This field theory is asymptotically free, and we assume the existence of a mass gap $m$. The renormalized field operator $\Phi(x)$ is an average of $U(x)$ over a region of size $b$, where $\Lambda^{-1}<b\ll m^{-1}$, where $\Lambda$ is an ultraviolet cutoff and $m$ is the mass of the fundamental excitation. 

Though $\Phi(x)$ is a complex $N\times N$ matrix, which is
not directly proportional to the unitary matrix $U(x)$, we have the equivalence
\beq
\Phi(x) \sim {\rm Z}(g_{0},\Lambda)^{-1/2}U(x), \label{phi} \nonumber
\eeq 
in the sense that
\beq
\frac{1}{N}\left \langle 0\vert {\rm Tr}\;\Phi(x) \Phi(0)^{\dagger} \vert 0\right\rangle
={\rm Z}(g_{0},\Lambda)^{-1}
\frac{1}{N}\left \langle 0\vert {\rm Tr}\;U(x)U(0)^{\dagger} \vert 0\right\rangle.
\label{renormalization}
\eeq
The renormalization factor 
${\rm Z}(g_{0}(\Lambda),\Lambda)$ goes to zero as $\Lambda\rightarrow \infty$ and the coupling
$g_{0}(\Lambda)$ runs so that the mass gap
$m(g_{0}(\Lambda), \Lambda)$ is independent of $\Lambda$.

The form factors may be combined into an expression for 
vacuum expectation values of products of 
$\Phi(x)$ and $\Phi(x)^{\dagger}$. We will use them to find an expression for the Wightman correlation function 
\beq
{\mathcal W}(x) = \frac{1}{N}\langle 0\vert {\rm Tr}\,\, \Phi(0) \Phi(x)^{\dagger} \vert 0\rangle\,.
\label{wightman}
\eeq 
There are other integrable models for which  
Wightman functions have been found with the form-factor 
bootstrap. These include the sinh-Gordon model
\cite{VG}, the scaling limit of the two-dimensional
Ising model \cite{Ising} (for which other methods yield the same results \cite{Ising1}), the 
$Z_{N}$ or clock model (a generalization of the Ising model to $N$ states) \cite{ZN}, \cite{A-N-1} and  affine-Toda models \cite{A-N-1}.

Form factors of ${\rm O}(4)\simeq {\rm SU}(2)\times {\rm SU}(2)$ sigma models were investigated in References
\cite{KW}. Form factors of the ${\rm SU}(N)$-invariant chiral Gross-Neveu model have been found in Reference \cite{BFK} for arbitrary $N$.

Correlation functions in lower-dimensional models are not of interest only for their own sake. They have applications in situations where integrability is broken by interactions. What motivated our investigations was such an application of the ${\rm SU}(N)\times {\rm SU}(N)$ principal chiral model to 
SU($N$) gauge theories in $2+1$ dimensions \cite{Latt2+1}.

In the next section, we present the $1/N$-expansion of the {\em S} matrix, summarize the results of 
{\bf I} and review (briefly) the Smirnov axioms. In Section 3, we find all the form factors as $N\rightarrow \infty$. All of these have simple poles in some relative 
rapidities. In Section 4, we apply our result to the large-$N$ Wightman function. In Section 5, we find the time-ordered correlation function can be expressed as a sum of Feynman diagrams with massive propagators and spherical topology. We summarize our results and present a few open problems in the last section.

\section{The $1/N$-Expansion and Form-Factor Axioms}
\setcounter{equation}{0}
\renewcommand{\theequation}{II.\arabic{equation}}

In this section we briefly review the basic features of the principal chiral 
model in the planar limit, and summarize the results of {\bf I}.

The {\em S} matrix of the elementary excitations of the principal chiral model 
\cite{AALS}, \cite{Wiegmann}, \cite{Pol-Wieg} is written in terms of the incoming rapidities
$\theta_{1}$ and 
$\theta_{2}$ (here $(p_{j})_{0}=m\cosh\theta_{j}$,
$(p_{j})_{1}=m\sinh \theta_{j}$, relates the momentum vector $p_{j}$ and rapidity $\theta_{j}$),
outgoing rapidities
$\theta_{1}^{\prime}$ and $\theta_{2}^{\prime}$
and rapidity difference $\theta=\vert \theta_{12}\vert =\vert \theta_{1}-\theta_{2}\vert $. The 
two-particle {\em S} matrix is
\begin{eqnarray}
S_{PP}(\theta)
=
\frac{\sin (\theta/2-\pi{\rm i}/N)}{\sin(\theta/2+\pi{\rm i}/N)}\;S_{\rm CGN}(\theta)_{L}\otimes 
S_{\rm CGN}(\theta)_{R} ,\label{s-matrix}
\end{eqnarray}
where $S_{\rm CGN}(\theta)_{L,R}$, for either the subscript L (left) or R (right), is the {\em S} matrix of two elementary excitations, each of which is a vector $N$-plet,  of the chiral Gross-Neveu 
model \cite{BKKW} :
\begin{eqnarray}
S_{\rm CGN}(\theta)\!\!=\!\!\frac{\Gamma({\rm i}\theta/2\pi+1)\Gamma(-{\rm i}\theta/2\pi-1/N) }{\Gamma({\rm i}\theta/2\pi+1-1/N) \Gamma(-{\rm i}\theta/2\pi)}
\left( {\mathbbm 1}-\frac{2\pi{\rm i}}{N\theta}P\right), \nonumber
\end{eqnarray}
and $P$ interchanges the colors of the two chiral-Gross-Neveu $N$-plet excitations. The {\em S} matrix matrix of more than two excitations is built out of the two-particle {\em S} matrix using the factorization property, {\em i.e.} the Yang-Baxter equation.

The $1/N$-expansion of the particle-particle {\em S} matrix (\ref{s-matrix}) is \cite{Wiegmann}
\beq
S_{PP}(\theta)^{c_{2}d_{2};c_{1}d_{1}}_{a_{1}b_{1};a_{2}b_{2}}
=\left[ 1+O(1/N^{2})\right]
\left[\delta^{c_{2}}_{a_{2}}\delta^{d_{2}}_{b_{2}}\delta^{c_{1}}_{a_{1}} \delta^{d_{1}}_{b_{1}}
-\frac{2\pi {\rm i}}{N\theta}\left(
\delta^{c_{2}}_{a_{1}}\delta^{d_{2}}_{b_{2}}\delta^{c_{1}}_{a_{2}} \delta^{d_{1}}_{b_{1}}
+\delta^{c_{2}}_{a_{2}}\delta^{d_{2}}_{b_{1}}\delta^{c_{1}}_{a_{1}} \delta^{d_{1}}_{b_{2}}
\right)
-\frac{4\pi^{2}}{N^{2}\theta^{2}}
\delta^{c_{2}}_{a_{1}}\delta^{d_{2}}_{b_{1}}\delta^{c_{1}}_{a_{2}} \delta^{d_{1}}_{b_{2}}
\right]. \label{expanded-s-matrix}
\eeq
The scattering matrix of one particle and one antiparticle $S_{PA}(\theta)$ is obtained by crossing
(\ref{expanded-s-matrix}) from the $s$-channel to the $t$-channel:
\beq
&S_{PA}\!\!\!&\!\!\!(\theta)^{d_{2}c_{2};c_{1}d_{1}}_{a_{1}b_{1};b_{2}a_{2}}=\left[ 1+O(1/N^{2})\right] \nonumber \\
&\times& \left[
\delta^{d_{2}}_{b_{2}}\delta^{c_{2}}_{a_{2}}\delta^{c_{1}}_{a_{1}}\delta^{d_{1}}_{b_{1}}
-\frac{2\pi{\rm i}}{N{\hat \theta}}
\!\left(\! \delta_{a_{1}a_{2}}\delta^{c_{1}c_{2}}\delta^{d_{2}}_{b_{2}}\delta^{d_{1}}_{b_{1}}
+\delta^{c_{2}}_{a_{2}}\delta^{c_{1}}_{a_{1}}\delta_{b_{1}b_{2}}\delta^{d_{1}d_{2}}
\! \right)\! -\frac{4\pi^{2}}{N^{2}{\hat \theta}^{2}}
\delta_{a_{1}a_{2}}\delta^{c_{1}c_{2}}\delta_{b_{1}b_{2}}\delta^{d_{1}d_{2}} 
  \right], \label{crossed-S}
\eeq 
where ${\hat \theta}=\pi {\rm i}-\theta$ is the crossed rapidity difference. As in {\bf I}, we define the generalized {\em S} matrix by replacing
$\theta=\vert\theta_{12}\vert$ with
$\theta=\theta_{12}$. 

The large-$N$ limit we consider is the standard 't Hooft limit. We assume the 
mass gap is fixed, as 
$N$ is  taken to infinity. 

There are bound states of the elementary particles, corresponding to poles of the 
{\em S} matrix. As we mentioned in {\bf I}, only one bound state plays a role in 
the correlation functions of 
$\Phi$. This bound state, namely the antiparticle,  consists of $N-1$ fundamental particles.

The first form factor (discussed in {\bf I}) is the simple normalization condition
\beq
\langle 0\vert \Phi(0)_{b_{0} a_{0}}\vert P,\theta, a_{1}, b_{1}\rangle
=N^{-1/2}\delta_{a_{0} a_{1}}\delta_{b_{0} b_{1}}, \label{norm}
\eeq
where the ket on the right is a one-particle state, with rapidity $\theta$ and left and right colors
$a_{1}$ and $b_{1}$, respectively.

The 
{\em S} matrix can be determined, assuming unitarity, factorization (the Yang-Baxter relation) and maximal analyticity. The excitations which survive in the $N\rightarrow \infty$ limit
have two color indices from $1$ to $N$. One can view these excitations as a bound pair of two quarks of different color sectors (or alternatively as a quark
in one color sector and an antiquark in the other).

The Zamolodchikov algebra is spanned by particle-creation 
operators ${\mathfrak A}^{\dagger}_{P}(\theta)_{ab}$ and
antiparticle-creation operators ${\mathfrak A}^{\dagger}_{A}(\theta)_{ba}$. These operators satisfy what 
is essentially a non-Abelian parastatistics relation:
\beq
{\mathfrak A}^{\dagger}_{P}(\theta_{1})_{a_{1}b_{1}}\,
{\mathfrak A}^{\dagger}_{P}(\theta_{2})_{a_{2}b_{2}}
&=&S_{PP}(\theta_{12})^{c_{2}d_{2};c_{1}d_{1}}_{a_{1}b_{1};a_{2}b_{2}}\;
{\mathfrak A}^{\dagger}_{P}(\theta_{2})_{c_{2}d_{2}}\,
{\mathfrak A}^{\dagger}_{P}(\theta_{1})_{c_{1}d_{1}} \nonumber \\
{\mathfrak A}^{\dagger}_{A}(\theta_{1})_{b_{1}a_{1}}\,
{\mathfrak A}^{\dagger}_{A}(\theta_{2})_{b_{2}a_{2}}
&=&S_{AA}(\theta_{12})^{d_{2}c_{2};d_{1}c_{1}}_{b_{1}a_{1};b_{2}a_{2}}\;
{\mathfrak A}^{\dagger}_{A}(\theta_{2})_{d_{2}c_{2}}\,
{\mathfrak A}^{\dagger}_{A}(\theta_{1})_{d_{1}c_{1}} \nonumber \\
{\mathfrak A}^{\dagger}_{P}(\theta_{1})_{a_{1}b_{1}}\,
{\mathfrak A}^{\dagger}_{A}(\theta_{2})_{b_{2}a_{2}}
&=&S_{PA}(\theta_{12})^{d_{2}c_{2};c_{1}d_{1}}_{a_{1}b_{1};b_{2}a_{2}}\;
{\mathfrak A}^{\dagger}_{A}(\theta_{2})_{d_{2}c_{2}}\,
{\mathfrak A}^{\dagger}_{P}(\theta_{1})_{c_{1}d_{2}}\,.
\label{creation}
\eeq
Consistency of this algebra implies the Yang-Baxter equation.

An in-state is defined as a product of creation operators in the order of increasing rapidity, from right to left, acting on the vacuum, {\em e.g.}
\beq
\vert P,\theta_{1},a_{1},b_{1};A,\theta_{2},b_{2},a_{2},\dots \rangle_{\rm in}
={\mathfrak A}^{\dagger}_{P}(\theta_{1})_{a_{1}b_{1}}
{\mathfrak A}^{\dagger}_{A}(\theta_{2})_{b_{2}a_{2}}\cdots \vert 0\rangle,\;\;{\rm where}\;
\theta_{1}>\theta_{2}>\cdots
\eeq
Similarly, an out-state is a product of creation operators in the order of decreasing rapidity, from right to left, acting on the vacuum.

The {\em S} matrix 
becomes the unit operator as $N\rightarrow \infty$. Hence the basic dynamical field is
\beq
M(x)=\int \frac{d\theta}{2\pi}\,\left[{\mathfrak A}_{P}(\theta)
e^{{\rm i}m x^{0}\cosh \theta-{\rm i}m 
x^{1}\sinh \theta} 
+{\mathfrak A}^{\dagger}_{A}(\theta)e^{-{\rm i}m x^{0}\cosh \theta+{\rm i}m x^{1}\sinh \theta} 
\right], \label{master}
\eeq
where ${\mathfrak A}_{P}$ is the destruction operator of a particle (which is the adjoint of the operator  ${\mathfrak A}^{\dagger}_{P}$). We can think of this $N\times N$ matrix-field operator as the 
master field, since its response to an external source is the same as that of a classical field (we can similarly regard the free field appearing in the large-$N$ limit of an O($N$)-symmetric model
as the master field). In {\bf I}, we pointed out that the form factors express  
the renormalized field $\Phi(x)$ in terms of the field $M(x)$.

The form factors are matrix elements between the vacuum and multi-particle in-states of the field
operator $\Phi$. The action of the global-symmetry transformation on $\Phi$ and the creation operators is
\beq
\Phi(x)\rightarrow V_{L}\Phi(x) V_{R}, \;\;
{\mathfrak A}^{\dagger}_{P}(\theta)\rightarrow V_{R}^{\dagger}
{\mathfrak A}^{\dagger}_{P}(\theta)V_{L}^{\dagger},\;\;
{\mathfrak A}^{\dagger}_{A}(\theta)\rightarrow V_{L}
{\mathfrak A}^{\dagger}_{P}(\theta)V_{R}. \label{transformation-laws}
\eeq
Thus we expect that, for large $N$, the condition
\beq
\langle 0\vert \Phi(0) \vert \Psi \rangle\neq 0,
\nonumber
\eeq
on an in-state $\vert \Psi\rangle$, which is an eigenstate of particle number, holds only if
$\vert \Psi\rangle$ contains $M$ particles and $M-1$ antiparticles, for some $M=1,2,\dots$ . In {\bf I} we found these matrix elements for $M=1$ (equation (\ref{norm}) above) and $M=2$:
\beq
\langle 0 \vert\, \Phi(0)_{b_{0}a_{0}}\, \vert A,\theta_{1}, b_{1},a_{1}; P,\theta_{2}, a_{2},b_{2};
P,\theta_{3}, a_{3},b_{3}\rangle_{\rm in} 
&=&
\langle 0\vert \, \Phi(0)_{b_{0}a_{0}}\,\, {\mathfrak A}^{\dagger}_{A}(\theta_{1})_{b_{1}a_{1}}
{\mathfrak A}^{\dagger}_{P}(\theta_{2})_{a_{2}b_{2}}
{\mathfrak A}^{\dagger}_{P}(\theta_{3})_{a_{3}b_{3}}\,\vert 0\rangle
\nonumber \\
=\frac{1}{N^{3/2}}F_{1}(\theta_{1}, \theta_{2}, \theta_{3}) 
\delta_{a_{0}a_{2}}\delta_{b_{0}b_{3}} \delta_{b_{1}b_{2}}\delta_{a_{1}a_{3}}
&+&\frac{1}{N^{3/2}}F_{2}(\theta_{1}, \theta_{2}, \theta_{3}) 
\delta_{a_{0}a_{3}}\delta_{b_{0}b_{2}}\delta_{a_{1}a_{2}}\delta_{b_{1}b_{3}}\nonumber \\
+\frac{1}{N^{3/2}}F_{3}(\theta_{1}, \theta_{2}, \theta_{3}) 
\delta_{a_{0}a_{2}}\delta_{b_{0}b_{2}}\delta_{a_{1}a_{3}}\delta_{b_{1}b_{3}}
&+&\frac{1}{N^{3/2}}F_{4}(\theta_{1}, \theta_{2}, \theta_{3}) 
\delta_{a_{0}a_{3}}\delta_{b_{0}b_{3}}\delta_{b_{1}b_{2}}\delta_{a_{1}a_{2}}, 
\label{F-form-factors}
\eeq
for $\theta_{1}>\theta_{2}>\theta_{3}$,
\beq
\langle 0 \vert\, \Phi(0)_{b_{0}a_{0}}\, \vert P,\theta_{1}, a_{1},b_{1}; A,\theta_{2}, b_{2},a_{2};
P,\theta_{3}, a_{3},b_{3}\rangle_{\rm in}  
&=&
\langle 0\vert \, \Phi(0)_{b_{0}a_{0}}\,\, {\mathfrak A}^{\dagger}_{P}(\theta_{2})_{a_{2}b_{2}}
{\mathfrak A}^{\dagger}_{A}(\theta_{1})_{b_{1}a_{1}}
{\mathfrak A}^{\dagger}_{P}(\theta_{3})_{a_{3}b_{3}}\,\vert 0\rangle
\nonumber \\
=\frac{1}{N^{3/2}}{\tilde F}_{1}(\theta_{1}, \theta_{2}, \theta_{3}) 
\delta_{a_{0}a_{2}}\delta_{b_{0}b_{3}} \delta_{b_{1}b_{2}}\delta_{a_{1}a_{3}}
&+&\frac{1}{N^{3/2}}{\tilde F}_{2}(\theta_{1}, \theta_{2}, \theta_{3}) 
\delta_{a_{0}a_{3}}\delta_{b_{0}b_{2}}\delta_{a_{1}a_{2}}\delta_{b_{1}b_{3}}\nonumber \\
+\frac{1}{N^{3/2}}{\tilde F}_{3}(\theta_{1}, \theta_{2}, \theta_{3}) 
\delta_{a_{0}a_{2}}\delta_{b_{0}b_{2}}\delta_{a_{1}a_{3}}\delta_{b_{1}b_{3}}
&+&\frac{1}{N^{3/2}}{\tilde F}_{4}(\theta_{1}, \theta_{2}, \theta_{3}) 
\delta_{a_{0}a_{3}}\delta_{b_{0}b_{3}} \delta_{b_{1}b_{2}}\delta_{a_{2}a_{1}},
\label{tilde-F}
\eeq
for $\theta_{2}>\theta_{1}>\theta_{3}$, and
\beq
\langle 0 \vert\, \Phi(0)_{b_{0}a_{0}}\, \vert P,\theta_{1}, a_{1},b_{1}; P,\theta_{2}, a_{2},b_{2};
A,\theta_{3}, b_{3},a_{3}\rangle_{\rm in}  
&=&
\langle 0\vert \, \Phi(0)_{b_{0}a_{0}}\,\, {\mathfrak A}^{\dagger}_{P}(\theta_{2})_{a_{2}b_{2}}
{\mathfrak A}^{\dagger}_{P}(\theta_{3})_{a_{3}b_{3}}
{\mathfrak A}^{\dagger}_{A}(\theta_{1})_{b_{1}a_{1}}\,\vert 0\rangle
\nonumber \\
=\frac{1}{N^{3/2}}{\tilde{\tilde F}}_{1}(\theta_{1}, \theta_{2}, \theta_{3}) 
\delta_{a_{0}a_{2}}\delta_{b_{0}b_{3}} \delta_{b_{1}b_{2}}\delta_{a_{1}a_{3}}
&+&\frac{1}{N^{3/2}}{\tilde{\tilde F}}_{2}(\theta_{1}, \theta_{2}, \theta_{3}) 
\delta_{a_{0}a_{3}}\delta_{b_{0}b_{2}}\delta_{a_{1}a_{2}}\delta_{b_{1}b_{3}}\nonumber \\
+\frac{1}{N^{3/2}}{\tilde{\tilde F}}_{3}(\theta_{1}, \theta_{2}, \theta_{3}) 
\delta_{a_{0}a_{2}}\delta_{b_{0}b_{2}}\delta_{a_{1}a_{3}}\delta_{b_{1}b_{3}}
&+&\frac{1}{N^{3/2}}{\tilde{\tilde F}}_{4}(\theta_{1}, \theta_{2}, \theta_{3}) 
\delta_{a_{0}a_{3}}\delta_{b_{0}b_{3}} \delta_{b_{1}b_{2}}\delta_{a_{2}a_{1}},
\label{tilde-tilde-F}
\eeq
for $\theta_{3}>\theta_{1}>\theta_{2}$. 
where 
\beq
F_{1}(\theta_{1},\theta_{2}, \theta_{3})& =&
-\frac{4\pi}{(\theta_{12}+\pi{\rm i})(\theta_{13}+\pi{\rm i}) }+O(1/N),\;\;\;
F_{2}(\theta_{1},\theta_{2}, \theta_{3}) \;=\;
-\frac{4\pi}{(\theta_{12}+\pi{\rm i})(\theta_{13}+\pi{\rm i})}+O(1/N)\;,\nonumber \\
F_{3}(\theta_{1},\theta_{2}, \theta_{3})&=&O(1/N),\;\;\;\;
F_{4}(\theta_{1},\theta_{2}, \theta_{3})\;=\;O(1/N),
\label{3part}
\eeq
and, to order $1/N^{0}$, the functions with one or two tildes are the same as those
in (\ref{3part}) except for phases. We should mention that the vanishing of $F_{3}$ and
$F_{4}$ is essential. If these quantities were not zero,  the double poles  would lead to diverging, unphysical
$S$ matrix elements through the LSZ reduction formula.

Here is a quick (but incomplete) summary of Smirnov's form-factor axioms  \cite{Smirnov} for arbitrary particle
states:

\vspace{5pt}

\noindent
{\em Scattering Axiom} (Watson's theorem). From the Zamolodchikov algebra (\ref{creation}),
\beq
\langle 0\vert \Phi(\!\!\!\!\!\!\!&\!\!\!\!0\!\!\!\!&\!\!\!\!\!\!\!)_{b_{0}a_{0}}\;
{\mathfrak A}^{\dagger}_{I_{1}}(\theta_{1})_{C_{1}} 
\cdots
{\mathfrak A}^{\dagger}_{I_{j}}(\theta_{j})_{C_{j}} 
{\mathfrak A}^{\dagger}_{I_{j+1}}(\theta_{j+1})_{C_{j+1}} \cdots\cdots
{\mathfrak A}^{\dagger}_{I_{M}}(\theta_{M})_{C_{M}} \vert 0\rangle \nonumber \\
&=&
S_{I_{j}I_{j+1}}(\theta_{j\; j+1})^{\!\!\!C^{\prime}_{j+1}C^{\prime}_{j}}_{\;\;\;C_{j}C_{j+1}}
\langle 0\vert \Phi(0)_{b_{0}a_{0}}\;{\mathfrak A}^{\dagger}_{I_{1}}(\theta_{1})_{C_{1}} 
\cdots
{\mathfrak A}^{\dagger}_{I^{\prime}_{j+1}}(\theta_{j+1})_{C^{\prime}_{j+1}}
{\mathfrak A}^{\dagger}_{I^{\prime}_{j}}(\theta_{j})_{C^{\prime}_{j}} \cdots
{\mathfrak A}^{\dagger}_{I_{M}}(\theta_{M})_{C_{M}} \vert 0\rangle 
, \label{watson}
\eeq
where $I_{k}$, $k=1,\dots,M$ is $P$ or $A$ (particle or antiparticle) and $C_{k}$ denotes a pair
of indices (which may be written $a_{k}b_{k}$, for $C_{k}=P$ and
 $b_{k}a_{k}$, for $C_{k}=A$) and similarly for the primed indices.

\vspace{5pt}

\noindent
{\em Periodicity Axiom}.
\beq
\langle 0\vert \Phi(0)_{b_{0}a_{0}}\;{\mathfrak A}^{\dagger}_{I_{1}}(\theta_{1})_{C_{1}} 
{\mathfrak A}^{\dagger}_{I_{2}}(\theta_{2})_{C_{2}} \cdots
{\mathfrak A}^{\dagger}_{I_{n}}(\theta_{n})_{C_{n}} \vert 0\rangle 
=\langle 0\vert \Phi(0)_{b_{0}a_{0}}\;
{\mathfrak A}^{\dagger}_{I_{n}}(\theta_{n}-2\pi{\rm i})_{C_{n}} \;
{\mathfrak A}^{\dagger}_{I_{1}}(\theta_{1})_{C_{1}} \cdots
{\mathfrak A}^{\dagger}_{I_{n-1}}(\theta_{M-1})_{C_{n-1}} \vert 0\rangle .
\label{periodicity}
\eeq

\vspace{5pt}

\noindent
{\em Annihilation-Pole Axiom}.  This is a recursive relation, which fixes the residues of the poles of the form 
factors. This axiom and the previous one
are special cases of a generalized crossing formula obtained in Reference \cite{BabKarow}. 
\beq
&{\rm  Res}\!\!&\!\!\vert_{\theta_{1n}=-\pi{\rm i}} \,
\langle \, 0 \,\vert\, \Phi(0)_{b_{0}a_{0}}\, 
{\mathfrak A}^{\dagger}_{I_{1}}(\theta_{1})_{C_{1}} 
{\mathfrak A}^{\dagger}_{I_{2}}(\theta_{2})_{C_{2}} \cdots
{\mathfrak A}^{\dagger}_{I_{n}}(\theta_{n})_{C_{n}}\vert 0\rangle  \nonumber \\
&=&-2{\rm i} \langle \, 0 \,\vert\, \Phi(0)_{b_{0}a_{0}}\, \vert 
{\mathfrak A}^{\dagger}_{I_{2}}(\theta_{2})_{C_{2}^{\prime}} 
{\mathfrak A}^{\dagger}_{I_{3}}(\theta_{2})_{C_{3}^{\prime}} \cdots
{\mathfrak A}^{\dagger}_{I_{n-1}}(\theta_{n-1})_{C_{n-1}^{\prime}}\vert 0\rangle \nonumber \\
&\times&
\left[  S_{I_{1}I_{2}}(\theta_{12})^{C_{2}^{\prime}D_{1}}_{\;\;C_{1}C_{2}}
 S_{I_{1}I_{3}}(\theta_{13})^{C_{3}^{\prime}D_{2}}_{\;\;D_{1}C_{3}}
\cdots S_{I_{1}I_{n-1}}(\theta_{1\;n-1})^{C_{n}C_{n-1}^{\prime}}_{\;\;D_{n-2}C_{n-1}}
-\delta^{C_{n}}_{\;\;C_{1}}
\delta^{C_{2}^{\prime}}_{\;\;C_{2}}\delta^{C_{3}^{\prime}}_{\;\;C_{3}}
\cdots \delta^{C_{n-1}^{\prime}}_{\;\;C_{n-1}}
\right], 
\label{a-p}
\eeq
If we assume the Lehmann-Symanzik-Zimmermann (LSZ) formula 
for the connected part of the {\em S} matrix with  $n-2$ external lines, then (\ref{a-p}) implies a similar LSZ formula with $n$ external lines. See {\bf I} for some discussion of the relation between this axiom and the reduction formula, in the context of the
large-$N$ limit of the principal chiral model. 

\vspace{5pt}

\noindent
{\em Lorentz-Invariance Axiom}. For the scalar operator $\Phi$, this takes the form
\beq
\langle 0\vert \Phi(0)_{b_{0}a_{0}}\;{\mathfrak A}^{\dagger}_{I_{1}}(\theta_{1}+\Delta \theta)_{C_{1}} 
\cdots
{\mathfrak A}^{\dagger}_{I_{M}}(\theta_{M}+\Delta \theta)_{C_{M}} \vert 0\rangle 
=
\langle 0\vert \Phi(0)_{b_{0}a_{0}}\;{\mathfrak A}^{\dagger}_{I_{1}}(\theta_{1})_{C_{1}} 
\cdots
{\mathfrak A}^{\dagger}_{I_{M}}(\theta_{M})_{C_{M}} \vert 0\rangle ,
\label{lorentz}
\eeq
for an arbitrary boost $\Delta \theta$.

\vspace{5pt}

\noindent
{\em Bound-State Axiom}. This axiom says that there are poles on the imaginary axis of 
rapidity differences
$\theta_{jk}$, due to bound 
states. we will not discuss it further, because
there are no bound states in the 't Hooft limit.

\vspace{5pt}

\noindent
{\em Minimality Axiom}. In general, form factors are holomorphic, except possibly for bound-state poles, for rapidity differences $\theta_{jk}$ in the complex strip $0<{\mathfrak Im}\;\theta_{jk}<2\pi$. The minimality axiom states that
form factors have as much analyticity as is consistent with the other axioms.

Some discussion of the meaning and use of these axioms in the context of the $N\rightarrow \infty$ limit
can be found in {\bf I}.

\section{Form factors for general in-states}
\setcounter{equation}{0}
\renewcommand{\theequation}{III.\arabic{equation}}

The general matrix element of $\Phi(0)$ between the vacuum and an $(M-1)$-antiparticle,
$M$-particle state has many terms. By comparing it to the {\em S}-matrix element describing the
scattering of $M$-particles, we can determine the most significant part of the form factors 
for large $N$. This part is proportional to $N^{-M+1/2}$. We denote left and right permutations (in the permutation group $S_{M}$) by $\sigma$ and $\tau$, respectively. We use the convention that $\sigma$ and $\tau$ take the set of numbers $0,\;1,\;2,\dots,\; M-1$ to 
$\sigma(0), \;\sigma(1),\;\dots,\; \sigma(M-1)$ and $\tau(0), \;\tau(1),\;\dots,\; \tau(M-1)$, respectively. The most general form factor of the renormalized field is
\beq
\!\!\!\!\!\!\!\langle 0\vert \Phi(0)_{b_{0}a_{0}}\;{\mathfrak A}^{\dagger}_{I_{1}}(\theta_{1})_{C_{1}} 
{\mathfrak A}^{\dagger}_{I_{2}}(\theta_{2})_{C_{2}} \cdots
{\mathfrak A}^{\dagger}_{I_{2M-1}}(\theta_{2M-1})_{C_{2M-1}} \vert 0\rangle 
=
\frac{\sqrt N}{N^{M}} \!\!\sum_{\sigma,\tau\in S_{M}}
F_{\sigma \tau}(\theta_{1},\theta_{2},\dots,\theta_{2M-1})
\!\prod_{j=0}^{M-1}\delta_{a_{j}\;a_{\sigma(j)+M}} 
\delta_{b_{j}\;b_{\tau(j)+M}} .
\label{M-FF}
\eeq
The order $(1/N)^{0}$ parts of the coefficients of the
tensors 
\beq
N^{-M+1/2}\prod_{j=0}^{M-1}\delta_{a_{j} \;a_{\sigma(j)+M}} \delta_{b_{j}\;b_{\tau(j)+M}}\,, \label{tensors}
\eeq
that is $F^{0}_{\sigma \tau}(\theta_{1},\theta_{2},\dots,\theta_{2M-1})$, are the same, no matter the order of the creation operators on the left-hand side of (\ref{M-FF}), except for a 
phase, as we explain below.

The function $F_{\sigma \tau}$ can be expanded in powers of $1/N$, {\em i.e.}
\beq
F_{\sigma \tau}(\theta_{1},\theta_{2},\dots,\theta_{2M-1})=
F^{0}_{\sigma \tau}(\theta_{1},\theta_{2},\dots,\theta_{2M-1})+
\frac{1}{N}F^{1}_{\sigma \tau}(\theta_{1},\theta_{2},\dots,\theta_{2M-1})+\cdots\;,
\label{M-FFexp}
\eeq
We only consider only the leading term on the right-hand side of (\ref{M-FFexp}) here.

Suppose we interchange two adjacent creation operators in the left-hand side of 
(\ref{M-FF}). The scattering axiom (\ref{watson}) implies that as $N\rightarrow \infty$:
\begin{enumerate}
\item If both creation operators create an antiparticle or both operators create a particle, the result is the interchange of the rapidities of these two operators, in the function $F^{0}_{\sigma\tau}$.
\item If one operator creates an antiparticle with rapidity $\theta_{j}$ and colors 
$a_{j}$, $b_{j}$ and 
the other operator creates a particle with rapidity $\theta_{k}$ and colors $a_{k}$, $b_{k}$,
and $\sigma(j)+M\neq k$, $\tau(j)+M\neq k$,
there is no effect on the function $F^{0}_{\sigma\tau}$.
\item   If one operator creates an antiparticle with rapidity $\theta_{j}$ and colors 
$a_{j}$, $b_{j}$ and
the other operator creates a particle with rapidity $\theta_{k}$ and colors $a_{k}$, $b_{k}$,
and  $\sigma(j)+M= k$, $\tau(j)+M\neq k$,
then $F^{0}_{\sigma\tau}$ is multiplied by the phase  
$\frac{\theta_{jk}+\pi {\rm i}}{\theta_{jk}-\pi{\rm i}}$.
\item   If one operator creates an antiparticle with rapidity $\theta_{j}$ and colors $a_{j}$, $b_{j}$ and
the other operator creates a particle with rapidity $\theta_{k}$ and colors $a_{k}$, $b_{k}$,
and  $\sigma(j)+M\neq k$, $\tau(j)+M= k$,
then $F^{0}_{\sigma\tau}$ is multiplied by the phase 
$\frac{\theta_{jk}+\pi {\rm i}}{\theta_{jk}-\pi{\rm i}}$.
\item   If one operator creates an antiparticle with rapidity $\theta_{j}$ and colors 
$a_{j}$, $b_{j}$ and the other operator creates a particle with rapidity $\theta_{k}$ and colors 
$a_{k}$, $b_{k}$, and  $\sigma(j)+M=k$, $\tau(j)+M= k$,
then $F^{0}_{\sigma\tau}$ is multiplied by the phase 
$\left(\frac{\theta_{jk}+\pi {\rm i}}{\theta_{jk}-\pi{\rm i}}\right)^{2}$.
\end{enumerate}

Statements 1. through 5. above are straightforward generalizations of the $M=2$ case, discussed in 
{\bf I}. The interchange of two creation operators has no effect at leading order in $1/N$, unless indices are contracted to make a factor of $N$. This factor of $N$  compensates for the terms of order $1/N$ in the {\em S} matrix. If no indices are contracted, the only part of the {\em S} matrix which contributes is unity. If the creation operators have an index in common, the phase $\frac{\theta_{jk}+\pi {\rm i}}{\theta_{jk}-\pi{\rm i}}$ appears, just as for the $M=2$ case. If two indices are contracted, this phase is squared. This is why Watson's theorem is meaningful as $N\rightarrow \infty$, despite there being no scattering!

Consider the following:
\beq
\langle \!\!\!\!\!&\!\!\!\!\!0\!\!\!\!\!\!&\!\!\!\! \vert \Phi(0)_{b_{0}a_{0}}\;
{\mathfrak A}^{\dagger}_{A}(\theta_{1})_{b_{1}a_{1}} 
\cdots \;{\mathfrak A}^{\dagger}_{A}(\theta_{M-1})_{b_{M-1}a_{M-1}}
{\mathfrak A}^{\dagger}_{P}(\theta_{M})_{a_{M}b_{M}}
\cdots
\;{\mathfrak A}^{\dagger}_{P}(\theta_{2M-1})_{a_{2M-1}b_{2M-1}} \vert \;0\;\rangle 
\nonumber \\
&=&
N^{-M+1/2}\sum_{\sigma,\tau\in S_{M}}
F_{\sigma \tau}(\theta_{1},\theta_{2},\dots,\theta_{2M-1})
\prod_{j=0}^{M-1}\delta_{a_{j}\;a_{\sigma(j)+M}} 
\delta_{b_{j}\;b_{\tau(j)+M}} ,
\label{special-FF}
\eeq
which is a special case of
(\ref{M-FF}). We interchange the leftmost creation operator
${\mathfrak A}^{\dagger}_{A}(\theta_{1})_{b_{1}a_{1}}$
consecutively with all the other creation operators. In other words, we are ``pushing"
${\mathfrak A}^{\dagger}_{A}(\theta_{1})_{b_{1}a_{1}}$ to the right, past all the other creation
operators. The periodicity axiom for $\theta_{1}$, together with the conditions 1. through 5., implies 
that the $N\rightarrow\infty$ limit of the function 
$F_{\sigma\tau}$ in (\ref{special-FF}) has the following structure, as a function of $\theta_{1}$
\beq
F^{0}_{\sigma \tau}(\theta_{1},\theta_{2},\dots,\theta_{2M-1})
\sim [\theta_{1}-\theta_{\sigma(1)+M}+\pi{\rm i}]^{-1} [\theta_{1}-\theta_{\tau(1)+M}+
\pi{\rm i}]^{-1}h(\theta_{1},\dots,\theta_{2M-1})  \nonumber
\eeq
where $h(\theta_{1},\dots, \theta_{2M-1})$ is some function which is analytic and periodic, with period $2\pi{\rm i}$, in $\theta_{1}$, 
$\theta_{\sigma(1)+M}$ and $\theta_{\tau(1)+M}$.

Now suppose we start with the same expression, namely (\ref{special-FF}), and interchange
${\mathfrak A}^{\dagger}_{A}(\theta_{1})_{b_{1}a_{1}}$ with 
${\mathfrak A}^{\dagger}_{A}(\theta_{2})_{b_{2}a_{2}}$.
This has the effect on the function $F_{\sigma\tau}$ of interchanging the arguments $\theta_{1}$ and
$\theta_{2}$.  Then we can apply a procedure similar to that of the last paragraph, pushing ${\mathfrak A}^{\dagger}_{A}(\theta_{2})_{b_{2}a_{2}}$ all the way to the right. We repeat this procedure for all the creation operators for antiparticles. We conclude that
$F^{0}_{\sigma\tau}$ has the form
\beq
F^{0}_{\sigma \tau}(\theta_{1},\theta_{2},\dots,\theta_{2M-1})
=
\frac{g_{\sigma \tau}(\theta_{1},\dots, \theta_{2M-1})}{\prod_{j=1}^{M-1} 
[\theta_{j}-\theta_{\sigma(j)+M}+\pi{\rm i}][\theta_{j}-\theta_{\tau(j)+M}+
\pi{\rm i}]}, \label{polesFF}
\eeq
where $g_{\sigma \tau}(\theta_{1},\dots, \theta_{2M-1})$ is holomorphic and periodic in each $\theta_{j}$, $j=1,\dots,2M-1$, with period $2\pi {\rm i}$. In (\ref{polesFF}), there are possible poles occurring
at $\theta_{j}-\theta_{\sigma(j)+M}=-\pi{\rm i}$
and
$\theta_{j}-\theta_{\tau(j)+M}=-\pi{\rm i}$. The maximum number of such poles on the right-hand side of
(\ref{polesFF}), including multiplicity, is $2M-2$. This is precisely the number needed to generate the  connected part of the {\em S} matrix (from the reduction formula) to leading order in $1/N$.

A choice of (\ref{polesFF}), satisfying the annihilation-pole axiom and having as much analyticity as possible is
\beq
F^{0}_{\sigma \tau}(\theta_{1},\theta_{2},\dots,\theta_{2M-1})
=
\frac{ (-4\pi)^{M-1}K_{\sigma \tau} }{\prod_{j=1}^{M-1} 
[\theta_{j}-\theta_{\sigma(j)+M}+\pi{\rm i}][\theta_{j}-\theta_{\tau(j)+M}+
\pi{\rm i}]},  \label{MFF}
\eeq
where 
\beq
K_{\sigma \tau}=\left\{ \begin{array}{cc} 1\;,& \;\sigma(j)\neq \tau(j), \;{\rm for \;all} \;j
\\
0\;,& \!\!\!\!\!\!\!\!\!\!\!\!\!\!\!\!\!\!\!\!\!\!\!\!\!\!\!\!\!\!\!\!\!\!\!\!{\rm otherwise} 
\end{array} \right.\;\;. \label{Kdef}
\eeq    
Notice that the expression for $K_{\sigma \tau}$ insures the absence of double poles. We 
recover (\ref{3part}) for $M=1$. Thus $K_{\sigma \tau}$ is unity if and only if the permutation
$\sigma o \tau^{-1}$ has no 
fixed points, {\em i.e.} has the smallest possible fundamental character in $S_{M}$. The number of pairs 
$\sigma$ and $\tau$ in $S_{M}$ satisfying this condition is $(M-1)!M!\,$. Together, (\ref{special-FF}), (\ref{MFF}) and (\ref{Kdef}) 
yield the form factors.

\section{Wightman Functions}
\setcounter{equation}{0}
\renewcommand{\theequation}{IV.\arabic{equation}}

The Wightman function is obtained from the form factors using the completeness of in-states:
\beq
{\mathcal W}(x)\; =\;\frac{1}{N}\sum_{a_{0},b_{0}}\sum_{X}\langle 0 \vert \Phi(0)_{b_{0}a_{0}} \vert X\rangle_{\rm in\;\;in}
\langle X \vert \;[\Phi(0)_{b_{0}a_{0}}]^{*} \;\vert 0\rangle e^{{\rm i}p_{X}\cdot x} 
\;=\;\frac{1}{N}\sum_{a_{0},b_{0}}\sum_{X}\,
\left\vert \langle 0 \vert \Phi(0)_{b_{0}a_{0}} \vert X\rangle_{\rm in}\right\vert^{2}\,
e^{{\rm i}p_{X}\cdot x} , \nonumber
\eeq
where $X$ denotes an arbitrary choice of particles, momenta and colors and where $p_{X}$ is the 
momentum eigenvalue of the state $\vert X\rangle$. From the result of the last section,
\beq
{\mathcal W}(x) = \sum_{M=1}^{\infty} \frac{1}{(M-1)!}\frac{1}{M!}\int \,\left( 
\prod_{j=1}^{2M-1}\frac{d\theta_{j}}{4\pi}\right)\,\sum_{\sigma \tau}\,
\frac{(4\pi)^{2M-2}\;K_{\sigma \tau}\;\exp({{\rm i}x\cdot \sum_{j=1}^{2M-1}p_{j}})
}{\prod_{j=1}^{M-1} 
\vert \theta_{j}-\theta_{\sigma(j)+M}+\pi{\rm i}\vert^{2}\vert \theta_{j}-\theta_{\tau(j)+M}+
\pi{\rm i}\vert^{2}}+O(1/N)\;, \label{tentative}
\eeq
where $p_{j}=m(\cosh\theta_{j},\sinh\theta_{j})$. Notice that the leading term is of order $1/N^{0}$. We did the sum over all color indices on the right-hand side of 
(\ref{tentative}), using
\beq
\sum_{a_{0},\dots,a_{2M-1},b_{0},\dots,b_{2M-1}}K_{\sigma\tau}K_{\omega \varphi}\left[\prod_{j=0}^{M-1}\delta_{a_{j}\;a_{\sigma(j)+M}} 
\delta_{b_{j}\;b_{\tau(j)+M}}\right]
\left[\prod_{j=0}^{M-1}\delta_{a_{j}\;a_{\omega(j)+M}} 
\delta_{b_{j}\;b_{\varphi(j)+M}}\right]
=N^{2M-2}K_{\sigma \tau}\left[\delta_{\sigma \omega}\delta_{\tau \varphi}+O(1/N)\right]\!\!, 
\label{tensorsum}
\eeq
for permutations $\sigma, \tau, \omega,\varphi\in S_{M}$. The $M=2$ case of  (\ref{tensorsum}) was discussed in {\bf I}. This relation tells us that the sum over the product of two of the color tensors (\ref{tensors}) will not contribute as $N\rightarrow \infty$, unless they are the same tensor. 

We can further simplify Equation (\ref{tentative}). Each contribution from a 
pair of permutations $\sigma$, $\tau$, satisfying $K_{\sigma\tau}=1$ on the right-hand side of 
(\ref{tentative}) is the same, after integrating
over the rapidities $\theta_{1}, \dots, \theta_{2M-1}$.  We can therefore pick one pair of permutations and
multiply by $M!(M-1)!$ (canceling a similar factor in the denominator). We choose the identity for $\sigma$ and a cyclic permutation for $\tau$:
\beq
\sigma(j)=j,\;\; \tau(j)=j+1\; ({\rm mod} \;M), \; {\rm for} \;j=0,\dots,M-1 \;. \nonumber
\eeq
The Wightman function is therefore
\beq
{\mathcal W}(x)=\int \frac{d\theta}{4\pi} e^{ {\rm i} x\cdot p}&+&
\frac{1}{4\pi}\sum_{M=2}^{\infty}\int d\theta_{1}\cdots d\theta_{2M-1}\;
e^{{\rm i}x\cdot \sum_{j=1}^{2M-1}p_{j}}
\frac{1}{(\theta_{1}-\theta_{M})^{2}+\pi^{2}}
\frac{1}{(\theta_{M}-\theta_{2})^{2}+\pi^{2}}
\frac{1}{(\theta_{2}-\theta_{M+1})^{2}+\pi^{2}} \nonumber \\
&\times& \frac{1}{(\theta_{M+1}-\theta_{3})^{2}+\pi^{2}}
\;\cdots\; 
\frac{1}{(\theta_{M-2}-\theta_{2M-1})^{2}+\pi^{2}}
\frac{1}{(\theta_{2M-1}-\theta_{M-1})^{2}+\pi^{2}}\;.
\nonumber
\eeq
The first two terms of this series were presented in {\bf I}. We relabel the indices on rapidities by 
\beq
\theta_{1}\rightarrow \theta_{1}, \;\theta_{M}\rightarrow \theta_{2}, \;\theta_{2}\rightarrow \theta_{3},\;\dots,\;
\theta_{2M-1}\rightarrow \theta_{2M-2},\; \theta_{M-1}\rightarrow \theta_{2M-1}, \nonumber
\eeq
obtaining
\beq
{\mathcal W}(x)=\int \frac{d\theta}{4\pi} e^{{\rm i}m(x^{-}e^{\theta}+x^{+}e^{-\theta})}+ \frac{1}{4\pi}\sum_{l=1}^{\infty}\int d\theta_{1}\cdots d\theta_{2l+1}
\exp\left[ {\rm i}\sum_{j=1}^{2l+1}m (x^{-}e^{\theta_{j}}+x^{+}e^{-\theta_{j}})\right]
\prod_{j=1}^{2l}\frac{1}{(\theta_{j}-\theta_{j+1})^{2}+\pi^{2}}
\;,
\label{series}
\eeq
where $l=M-1$ and $x^{\pm}=(x^{0}\pm x^{1})/2$ are light-cone coordinates.

The terms in the series (\ref{series}) are multiple integrals over the  
Poisson kernel for the upper half-plane. Suppose that $f(\theta)$ is a function of real $\theta$, such that $\vert f(\theta)\vert\le C$, for some real positive constant $C$. The Poisson kernel integrated over $f$ is
\beq
Pf(\theta,y)=\frac{1}{\pi}\int_{-\infty}^{\infty} d\theta^{\prime} \frac{yf(\theta^{\prime})}{(\theta-\theta^{\prime})^{2}+y^{2}} ,\;\;y\ge 0\;.\nonumber
\eeq 
This function is harmonic everywhere in the upper half-plane, with the Dirichlet boundary condition 
$Pf(\theta,0)=f(\theta)$. The terms in (\ref{series}) are repeated integrations 
of this type, with $y=\pi$. This guarantees that each term in the series is 
finite (though it does not guarantee convergence of the series). The subtlety 
in evaluating the terms in the series 
(\ref{series}) is that
$\exp {\rm i} m(x^{-}e^{\theta_{j}}+x^{+}e^{-\theta_{j}})$, an analytic function of complex $\theta_{j}$, is not bounded in the upper half-plane as $\vert \theta_{j}\vert \rightarrow\infty$. In particular, as 
${\mathfrak Re}\,\theta_{j}\rightarrow \pm \infty$, it diverges for some choices of 
${\mathfrak Im}\,\theta_{j}$.The integral over the Poisson
kernel, however, is bounded and harmonic, but not analytic.


\section{Time-Ordered Green's Functions}
\setcounter{equation}{0}
\renewcommand{\theequation}{V.\arabic{equation}}

If we time-order the fields of the correlation function of the last section, we replace the time coordinate 
$x^{0}$ by $\vert x^{0}\vert$. The Lorentz-invariant two-point Green's function: 
\beq
G(x)=\frac{1}{N} \langle 0 \vert  \;T \,{\rm Tr}\;\Phi(0) \Phi(x) \vert 0\rangle , \nonumber
\eeq
can be written as a sum of integrals over energy-momentum two-vectors, $p_{1},\;p_{2},\;\dots,\;p_{2l+1}$:
\beq
G(x)&=&
\sum_{l=0}^{\infty}\frac{1}{4\pi}\int d^{2}p_{1}\,d^{2}p_{2}\cdots d^{2}p_{2l+1} \;e^{-{\rm i} ( p_{1}+p_{2}+\cdots +p_{2l+1} )\cdot  x}
 \left[\left.\prod_{j=1}^{2l}\right\vert_{j\;{\rm even}}\;
\frac{1}{(\cosh^{-1}\frac{p_{j}\cdot p_{j+1}}{m^{2}})^{2}+\pi^{2}} \right]
 \nonumber \\
&\times& \frac{\rm i}{\pi(p_{1}^{2}-m^{2}+{\rm i}\epsilon)}\frac{\rm i}{\pi(p_{2}^{2}-m^{2}+{\rm i}\epsilon)}\cdots
\frac{\rm i}{\pi(p_{2l+1}^{2}-m^{2}+{\rm i}\epsilon)} \left[\left.\prod_{j=1}^{2l}\right\vert_{j \;{\rm odd}}\;
\frac{1}{(\cosh^{-1}\frac{p_{j}\cdot p_{j+1}}{m^{2}})^{2}+\pi^{2}} \right] , \label{GF}
\eeq
where $p_{j}\cdot p_{k}=\eta^{\mu\nu}p_{j\;\mu}p_{k\;\nu}$ and $p_{j}^{2}=p_{j}\cdot p_{j}$. Each finite term in this amplitude is given by a rainbow-type Feynman diagram, with two vertices and $2l$ 
loops. For $l=3$, this diagram is 
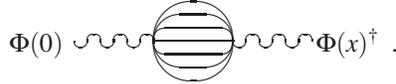
\begin{center}

\vspace{5pt}
\begin{picture}(150,20)(-10,0)

\put(64.5,10){\oval(30,30)}
\put(64.5,10){\oval(30,20)}
\put(64.5,10){\oval(30,10)}
\put(50,10){\line(1,0){30}}

\put(22,10){\oval(5,5)[b]}
\put(27,10){\oval(5,5)[t]}
\put(32,10){\oval(5,5)[b]}
\put(37,10){\oval(5,5)[t]}
\put(42,10){\oval(5,5)[b]}
\put(47,10){\oval(5,5)[t]}

\put(82,10){\oval(5,5)[b]}
\put(87,10){\oval(5,5)[t]}
\put(92,10){\oval(5,5)[b]}
\put(97,10){\oval(5,5)[t]}
\put(102,10){\oval(5,5)[b]}
\put(107,10){\oval(5,5)[t]}

\put(-5,6){$\Phi(0)$}
\put(111,6){$\Phi(x)^{\dagger}\;\;$.}

\end{picture}

\end{center}

\noindent
Each vertex is of order $2l+1$ (in other words, is joined by
$2l+1$ propagators). The massive propagators in (\ref{GF})
are those of the $l+1$ particles and $l$ antiparticles joining the two vertices. Though such diagrams, for $l>0$, are one-particle irreducible, (\ref{GF}) is the connected Green's function, not the one-particle-irreducible Green's function.

\section{Discussion}
\setcounter{equation}{0}
\renewcommand{\theequation}{VI.\arabic{equation}}

In this paper, we extended the derivation of the one- and three-excitation form factors of the renormalized field in {\bf I} to all the form factors of the this field. Using these form factors, we found expressions for correlation functions. 

It is important to know the behavior of the two-point Wightman function 
at short distances. Its Fourier transform, as a function of momentum $q$, must be  
consistent with  
asymptotic 
freedom. In particular, this function should be $\sim \sqrt{\log \vert q^{2}}\vert/q^{2}$, for large $q$. We hope 
to check that this behavior follows from (\ref{series}). If this can be 
done, it seems feasible to find coefficients of 
operator-product 
expansions. For example, we expect that for small $x$, 
\beq
\Phi(0)\Phi(x)^{\dagger} \simeq\; {\mathcal W}(x)[{\mathbbm 1}
+x^{\mu} U(0)\partial_{\mu}U(0)^{\dagger}]+\cdots 
\;=\;{\mathcal W}(x)[{\mathbbm 1}\;
+\;{\rm i}x^{\mu}{j^{\rm L}}_{\!\!\mu}(0)]+\cdots \;. 
\label{OPE}
\eeq
where ${j^{\rm L}}_{\!\!\mu}(x)=-{\rm i} U(x)\partial_{\mu}U(x)^{\dagger}$ is the left-handed current. The normalization of the second term in 
(\ref{OPE}) should be consistent with the $SU(\infty)$ current algebra. The completeness of in-states makes it possible to check that the form factors (\ref{MFF}), 
(\ref{Kdef})  are consistent with the form factors  
of current operators \cite{axel}. The latter form factors should be useful in the study of the large-$N$ limit of ${\rm SU}(N)$ gauge theories in $2+1$ dimensions, along the lines of
References \cite{Latt2+1}.

\begin{acknowledgments}
I thank Axel Cort\'es Cubero and Gerald Dunne for discussions. This work was supported in 
part by the National Science Foundation, under Grant No. 
PHY0855387, and by a grant from the PSC-CUNY. 
\end{acknowledgments}

\end{document}